\documentclass[
reprint,
superscriptaddress,
 amsmath,amssymb,
aps,
]{revtex4-2}

\usepackage{color}
\usepackage{graphicx}
\usepackage{dcolumn}
\usepackage{bm}
\usepackage{float}

\begin{document}


\title{Chaotic magnetization dynamics in magnetic Duffing oscillator}

\author{Ryo Tatsumi}
 \affiliation{Department of Applied Physics, Graduate School of Engineering, Tohoku University, Sendai, Miyagi 980-8579, Japan}
\author{Takahiro Chiba}
 \affiliation{Department of Applied Physics, Graduate School of Engineering, Tohoku University, Sendai, Miyagi 980-8579, Japan}
 \affiliation{Frontier Research Institute for Interdisciplinary Sciences, Tohoku University, Sendai, Miyagi 980-8578, Japan}
 \affiliation{Department of Information Science and Technology, Graduate School of Science and Engineering, Yamagata University, Yonezawa, Yamagata 992-8510, Japan}
 \author{Takashi Komine}
 \affiliation{Graduate School of Science and Engineering, Ibaraki University, Hitachi, Ibaraki, 316-8511, Japan}
\author{Hiroaki Matsueda}
 \affiliation{Department of Applied Physics, Graduate School of Engineering, Tohoku University, Sendai, Miyagi 980-8579, Japan}
 \affiliation{Center for Science and Innovation in Spintronics, Tohoku University, Sendai 980-8577, Japan}


\date{\today}
 
\begin{abstract}
We propose a magnetic analogy of the Duffing oscillator–-magnetic Duffing oscillator–-which is characterized by a double-well magnetic potential of a ferromagnet with a uniaxial magnetic anisotropy. 
Based on the linear stability analysis of the Landau–Lifshitz–Gilbert equation, we show that an external magnetic field applied perpendicular to the magnetic anisotropy field creates an anharmonicity on the magnetic potential, generating homoclinic orbits in the phase space. 
By evaluating the Lyapunov exponent, we demonstrate that the magnetic Duffing oscillator exhibits chaotic behaviors in the presence of periodically oscillating external forces: Oersted field and spin-orbit torque by considering the ferromagnet/heavy-metal bilayer. 
We also show that the external magnetic field can be adjusted to generate or modify homoclinic orbits, thereby controlling the parameter range of the oscillating external forces that induce chaos.
This work deepens our understanding of chaotic magnetization dynamics by bridging the fields of nonlinear dynamics and spintronics.
\end{abstract}


\maketitle

\section{Introduction}
Recently, nonlinear dynamics in physical systems have garnered much attention as a computational resource for chaotic neural network \cite{Aihira90,Yamada93,Pan21} and physical reservoir computing \cite{Tanaka19,Paquot12,Torrejon17}. 
Chaotic dynamics is essential to construct the architecture of the chaotic neural network that is applicable to associative memory and combinatorial optimization. 
It has also been reported that reservoir computing sometimes shows high-performance at the so-called ``edge of chaos'' which is a transient state between the periodic and chaotic states \cite{Legenstein07,Boedecker12,Bertschinger04,Yamaguchi23}. 
Thus, revealing the nonlinear dynamics in physical systems, including the chaotic state, is of great interest for neural network technologies. 

Among various physical phenomena exhibiting nonlinear dynamics, magnetization dynamics on the basis of the Landau–Lifshitz–Gilbert (LLG) equation has been studied actively because of the strong nonlinearity such as the Gilbert damping, spin-transfer torque, and magnetic anisotropy \cite{Bertotti09}. 
Spintronic devices, in particular, are considered promising candidates to achieve physical reservoir computing with high-performance and low-power energy consumption \cite{Contreras-Celada22,Taniguchi22}. 
In fact, the physical reservoir computing based on a spin-torque oscillator (STO) was experimentally demonstrated with a high score in the voice recognition task \cite{Torrejon17}. 
Also, numerous studies have reported chaotic magnetization dynamics in spintronics devices \cite{Li06,Yamaguchi19,Taniguchi19,Contreras-Celada22,Taniguchi22,Bragard11} especially in practical situations, in order to find the ``edge of chaos'' and enhance their potential as computational devices.
Li $\textit{et al}$. studied that chaos is induced by the specific shape of the magnetic potential due to the external field and demagnetization field in the spin-valve based STO in the presence of an alternating current \cite{Li06}.
Yamaguchi $\textit{et al}$. found that chaos occurs when the phase locking of STO in the magnetic tunnel junction is released by a microwave field or an alternating current \cite{Yamaguchi19}.
It is also known that the STO with a feedback input \cite{Taniguchi19} and the nano-oscillator driven by the voltage-controlled magnetic anisotropy (VCMA) effect \cite{Contreras-Celada22,Taniguchi22} exhibit chaotic magnetization dynamics.
However, mathematical analysis and identifying the origin of chaos in practical situations are challenging due to various nonlinear effects which mask our intuitive comprehension of chaos in the spintronic devices.

The Duffing oscillator \cite{Kovacic11}, one of the famous examples of chaos, describes the dynamics of a mass point in a double-well potential under an external periodic force and has a specific type of orbit known as a homoclinic orbit in the phase space. 
It is also known that the horseshoe chaos can appear when there is the homoclinic orbit in the phase space \cite{Wiggins03}.
The mechanism of chaos in mass point dynamics within the Euclidean phase space is well understood. 
In contrast, chaos in the classical mechanics of a spin including magnetization dynamics remains unclear due to the geometry of the phase space $S^2$. 
Unlike typical Hamiltonian systems, such as the Duffing oscillator, the classical mechanics of a spin cannot be described in terms of a global system of coordinates and momenta \cite{Altland06}. 
In other words, we cannot define the canonical variable of the spin globally in the phase space $S^2$ due to the existence of the singular point.
The most important difference is that the Lagrangian of the classical dynamics of a spin does not involve an ordinal kinetic term but instead has the spin Berry phase term which generates an emergent massless point particle on $S^2$ coupled to a monopole field.
Thus, the difference in geometry of the phase space and associated dynamics severely hinders making a simple analogy of chaos in the classical mechanics of a mass point.

In this paper, to understand the principle of chaos in the classical mechanics of a spin, we propose a ``toy-model'' of chaos, which is characterized by a double-well magnetic potential of a ferromagnet with a uniaxial magnetic anisotropy.
Based on the linear stability analysis of the LLG equation, we show that an anharmonicity is created on the magnetic potential by perpendicularly applying an external magnetic field to the direction of the uniaxial magnetic anisotropy, generating a homoclinic orbit in the phase space. 
Then, since the phase portrait of the LLG equation has a similar topology to the Duffing oscillator, we call the proposed model ``magnetic Duffing oscillator''.
By calculating the bifurcation diagram, the Poincar\'{e} section, and the Lyapunov exponent \cite{Pikovsky16,Shimada79}, we numerically demonstrate that the magnetic Duffing oscillator exhibits chaos in the presence of periodically oscillating external forces: Oersted field and spin-orbit torque (SOT) \cite{Manchon19}.
It is also confirmed that the chaotic dynamics disappears when the homoclinic orbit vanishes in the phase space. 
We find that by adjusting the external magnetic field, we can control the onset of chaos and the parameter range of the external oscillating forces that induce chaotic dynamics.
Since the main source of chaos in this model is the magnetic potential rather than the driving force, our model can be implemented with various types of spintronics devices, such as the ferromagnetic resonance (FMR), spin-torque ferromagnetic resonance (ST-FMR), STO, and VCMA oscillator.
Furthermore, we reveal the common structure underlying the proposed model and the Duffing oscillator using the phase portrait and Poincar\'{e} section, which implies that various phenomena observed in the Duffing equation, such as hyperchaos \cite{Kapitaniak93,Kingston22} and noise-induced transitions \cite{Aumaitre07}, may also appear in spintronics models.

The paper is organized as follows. In Sec.~\ref{model}, we describe the magnetic Duffing oscillator based on the LLG equation. 
In Sec.~\ref{Linear_stability}, by using the linear stability analysis of the LLG equation, we investigate the nature of the magnetic Duffing oscillator and show the condition under which the homoclinic orbit appears. In Sec.~\ref{Numerical analysis}, by using the bifurcation diagram, the Poincar\'{e} section, and the Lyapunov exponent, we demonstrate the emergence of chaos in the presence of the Oersted field and SOT induced by an alternating current.
We also associate the existence of the homoclinic orbit with the emergence of chaos. In Sec.~\ref{Discussion}, we discuss the applicability of the proposed model, its potential experimental realization, and previous studies that have explored the correspondence between the Duffing oscillator and spintronics devices.
A concluding summary is given in Sec.~\ref{Conclusion}

\section{Model}\label{model}
\begin{figure}[ptb]
\begin{centering}
\includegraphics[width=0.47\textwidth,angle=0]{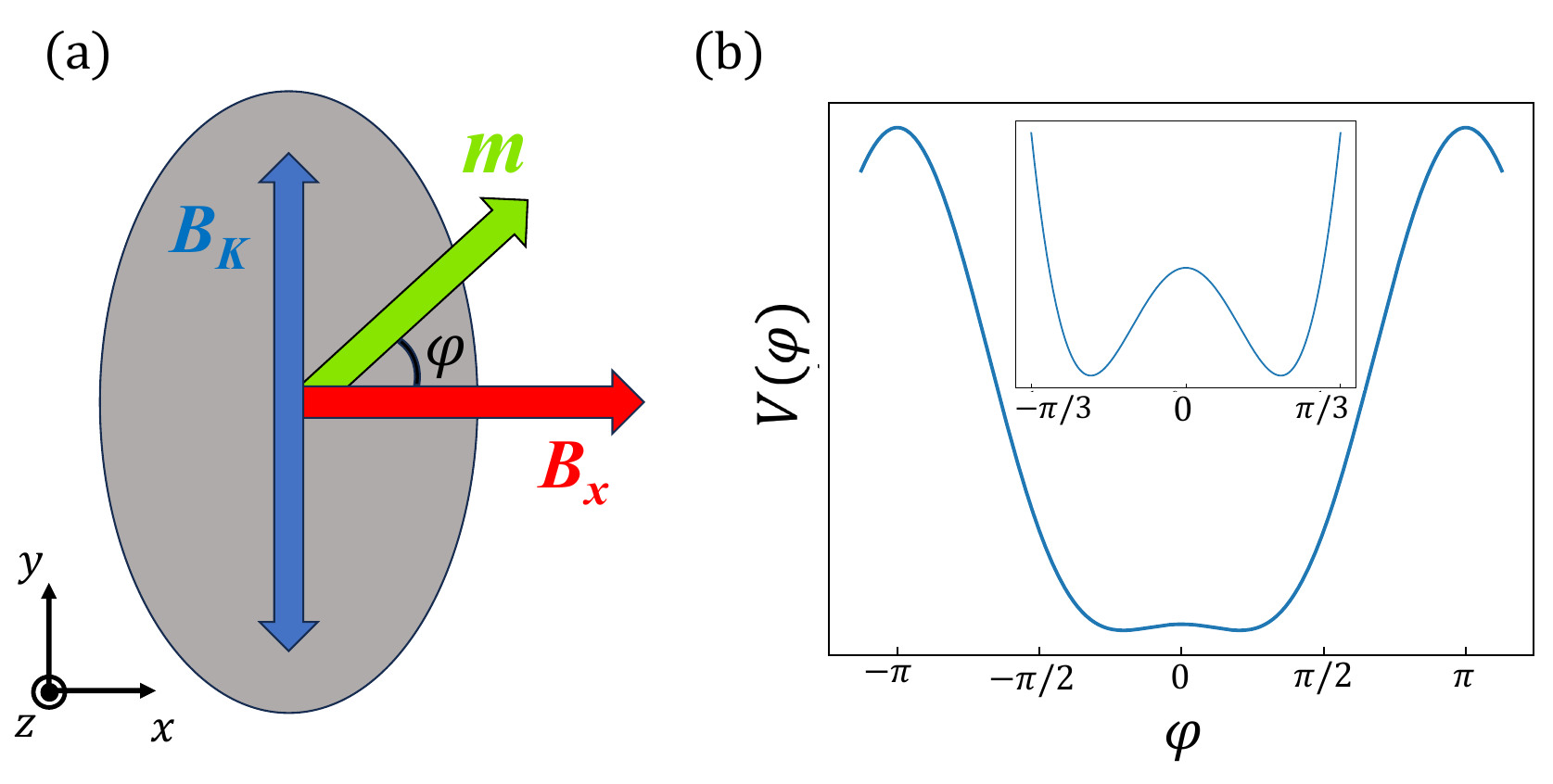} 
\par\end{centering}
\caption{
(a) Schematic illustration of the magnetic Duffing oscillator. An external magnetic field $B_x$ is applied along the $x$ axis and the uniaxial magnetic anisotropy field $B_K$ is along the $y$ axis. The azimuth of magnetization vector $\bm{m}$ on the $x$-$y$ plane is denoted by $\varphi$. Note that the magnetization has $z$ component $m_z$. (b) The potential shape of the magnetic Duffing oscillator in the $x$-$y$ plane for $B_x = 160$~mT and $B_K = 200$~mT.}
\label{fig:Magnetic Duffing Oscillator}
\end{figure}
Here, we introduce the magnetic Duffing oscillator consisting of a ferromagnet with a uniaxial magnetic anisotropy under an external magnetic field, as shown in Fig.~\ref{fig:Magnetic Duffing Oscillator}. Magnetization dynamics of a uniform ferromagnet is described by the LLG equation
\begin{equation}
\begin{split}
    \frac{d \bm{m}}{dt} = - \gamma \bm{m} & \times \bm{B}_{\mathrm{eff}} + \alpha \bm{m} \times \frac{d \bm{m}}{dt},
\end{split}
    \label{LLG equation}
\end{equation}
where $\bm{m}$, $\gamma$, $\bm{B}_{\mathrm{eff}}$, and $\alpha$ are the normalized magnetization vector by a saturation magnetization $M_{\rm s}$, the gyromagnetic ratio, the effective magnetic field, and the Gilbert damping constant, respectively.
The effective magnetic field $\bm{B}_{\mathrm{eff}}$ consists of
\begin{equation}
\begin{split}
    &\bm{B}_{\mathrm{eff}} = B_x \hat{\bm{e}}_x + B_K m_y \hat{\bm{e}}_y,
\end{split}
\label{magnetic_field}
\end{equation}
where $\{\hat{\bm{e}}_x,\hat{\bm{e}}_y, \hat{\bm{e}}_z\}$ are the unit vectors along the respective Cartesian axes, $B_x$ is a static external magnetic field, and $B_K$ is a uniaxial magnetic anisotropy field. Both $B_x$ and $B_K$ are positive. Note that $B_K$ is assumed to be a magnetocrystalline anisotropy or shape magnetic anisotropy. 

In the polar coordinates describing the Bloch sphere of ${\bm m}$, the magnetic potential is given by
\begin{equation}
\begin{split}
    V(\theta,\varphi) &= - \gamma \int d\bm{m} \cdot \bm{B}_{\mathrm{eff}}\\
    & = - \gamma \left(  B_x \sin{\theta} \cos{\varphi} + \frac{1}{2} B_K \sin^2{\theta} \sin^2{\varphi}\right),
\end{split}
\label{magnetic_potential}
\end{equation}
where $\theta$ and $\varphi$ are the polar and azimuth angles of the Bloch sphere, respectively. 
Then, the magnetization $\bm{m}$ is represented by $(m_x, m_y, m_z) = (\sin{\theta} \cos{\varphi}, \sin{\theta} \sin{\varphi}, \cos{\theta})$.
As shown in Fig.~\ref{fig:Magnetic Duffing Oscillator}(a), the external magnetic field is applied perpendicularly to the magnetic anisotropy field. When the strength of the external magnetic field is lower than that of the magnetic anisotropy field, the shape of the magnetic potential becomes a double-well, as shown in Fig.~\ref{fig:Magnetic Duffing Oscillator}(b). 

\section{Linear stability analysis}\label{Linear_stability}
\begin{figure}[ptb]
\begin{centering}
\includegraphics[width=0.47\textwidth,angle=0]{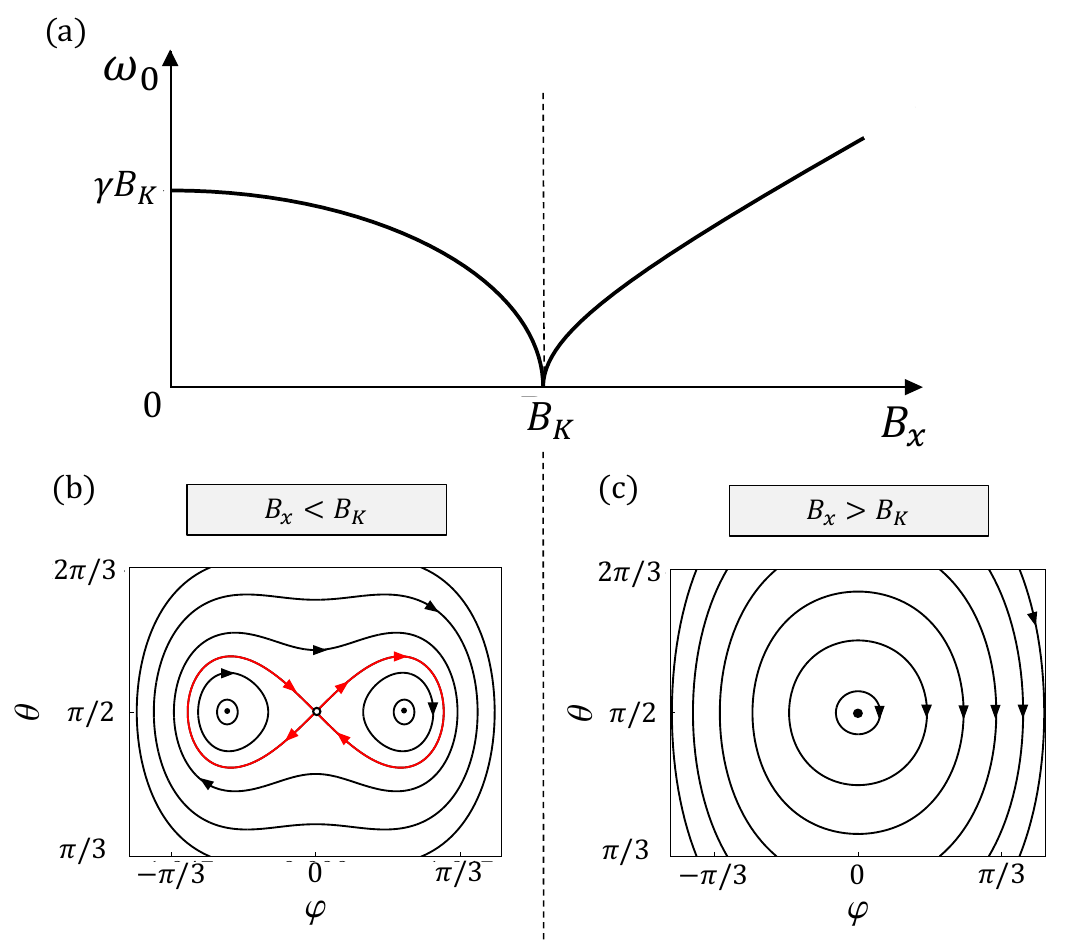} 
\par\end{centering}
\caption{
(a) The resonance (angular) frequency of the magnetic Duffing oscillator, which is obtained by the imaginary part of the eigenvalue of the Jacobian matrix at the equilibrium points in the phase space, as a function of the external magnetic field $B_x$.
Phase portraits excluding the relaxation by $\alpha = 0$ in the case of (b) $B_x < B_K$ and (c) $B_x > B_K$. 
Filled circles represent the center and an open circle represents the saddle. The red line is the homoclinic orbit, which is the trajectories that start from and end at the saddle point.}
\label{fig:Phase portrait}
\end{figure}

In this section, based on the linear stability analysis of the LLG equation, we analyze the nature of equilibrium points in the magnetic Duffing oscillator for various parameters and reveal the condition for the existence of the homoclinic orbit. 
To this end, we rewrite Eq.~(\ref{LLG equation}) in the polar coordinates excluding the Gilbert damping term by setting $\alpha = 0$. It is noted that due to the constraint $|\bm{m}| = 1$ the magnetization dynamics is confined on the Bloch sphere which corresponds to the phase space $S^2$. Hence,
\begin{equation}
\begin{split}
    \frac{d }{dt} \begin{pmatrix}
        \theta \\
        \varphi \\
    \end{pmatrix} &= \begin{pmatrix}
        \gamma B_\varphi(\theta, \varphi) \\
        - \frac{1}{\sin{\theta}}\gamma B_\theta(\theta, \varphi) \\
    \end{pmatrix},
\end{split}
\label{polar_LL}
\end{equation}
where the $B_\theta$ and $B_\varphi$ represent the $\theta$ and $\varphi$ components of the magnetic field, respectively. These components are given by
\begin{equation}
    \begin{split}
        B_\theta(\theta,\varphi) &= B_x \cos{\theta} \cos{\varphi} + B_K \sin{\theta} \cos{\theta} \sin{\varphi}^2,\\
        B_\varphi(\theta,\varphi) &= - B_x \sin{\varphi} + B_K \sin{\theta} \sin{\varphi} \cos{\varphi}.
    \end{split}
    \label{magnetic_field_sphe}
\end{equation}
Since Eq.~(\ref{polar_LL}) is a first-order differential equation for $\theta$ and $\varphi$, $\dot{\theta}$ and $\dot{\varphi}$ are defined at each point in the phase space of $\theta$ and $\varphi$. 
Therefore, the magnetization dynamics is described by the vector fields $\dot{\bm{x}} = \bm{f}(\bm{x})$ with $\bm{x} = (\theta,\varphi)$ in the phase space of $\theta$ and $\varphi$. 
Note that the imaginary part of the eigenvalues of the Jacobian matrix $J = \left. \partial \bm{f}/ \partial \bm{x} \right|_{\bm{x}=\bm{x}^*} $, where $\bm{x}^*$ represents an equilibrium point, corresponds to the angular velocity around these points.
In other words, if the eigenvalues of the Jacobian matrix are complex, such that $\lambda = a + i\omega_0 \mid a,\omega_0 \in \mathbb{R}$, the system undergoes rotation around the equilibrium points with an angular velocity $\omega = \omega_0$. 
Therefore, the resonance frequency of the magnetization is given by the imaginary part of the Jacobian matrix's eigenvalues, which is shown in Fig.~\ref{fig:Phase portrait}(a).
Furthermore, the sign of the real part of the eigenvalue $a$ determines the stability of the equilibrium point.

In the case of $B_x < B_K$, the equilibrium points $(\theta^*, \varphi^*) = (\pi/2,0)$ and $(\pi/2,\pm \varphi_0)$ satisfy $\partial \theta / \partial t = \partial \varphi / \partial t= 0$, where $\cos{\varphi_0} = B_x/B_K$. Then, the Jacobian matrix at $(\theta^*, \varphi^*) = (\pi/2,0)$ is given by
\begin{equation}
    J = \gamma
\begin{pmatrix}
0 & -B_x + B_K \\
B_x & 0 \\
\end{pmatrix}
.\label{J}
\end{equation}
Accordingly, the eigenvalues of this matrix are $\lambda_{1,2} = \pm \gamma \sqrt{B_x(B_K-B_x)}$, that is, $\lambda_1 < 0 < \lambda_2$ and hence the equilibrium point $(\theta^*, \varphi^*) = (\pi/2,0)$ is a saddle point. Next, the Jacobian matrix at $(\theta^*, \varphi^*) = (\pi/2,\pm \varphi_0)$ is given by 
\begin{equation}
    J = \gamma
\begin{pmatrix}
0 & \frac{B_x^2}{B_K} - B_K \\
B_K & 0 \\
\end{pmatrix}
.
\end{equation}
Accordingly, the eigenvalues of this matrix are $\lambda_{1,2} = \pm i \gamma \sqrt{B_K^2 - B_x^2}$, purely imaginary, and are thus centers. 
As shown in Fig.~\ref{fig:Phase portrait}(b), each stable center is surrounded by a family of small closed orbits. 
There are also large closed orbits that encircle all three fixed points. 
Additionally, the phase portraits have two special trajectories: these are the trajectories that start from and end at the same saddle point, which is the so-called homoclinic orbit.
The magnetic Duffing oscillator is expected to exhibit chaos because there are homoclinic orbits in the phase space \cite{Wiggins03}. 
Additionally, as shown in Fig.~\ref{fig:Phase portrait}(b), the two centers and the saddle point are aligned along the $\varphi$ direction, and two distinct homoclinic orbits originating from the same saddle point enclose each center in a butterfly-like shape.
Thus, the topology of the phase portrait is identical to that of the Duffing oscillator. This fact implies that the qualitative properties of magnetization dynamics are equivalent to those of the Duffing oscillator. Furthermore, as will be discussed in Sec.~\ref{Numerical analysis}, the trajectories, Poincaré section, and routes to chaos exhibit similar behavior.

On the other hand, in the case of $B_x > B_K$, the equilibrium point is only $(\theta^*, \varphi^*) = (\pi/2,0)$. The Jacobian matrix at this point is the same as Eq.~(\ref{J}). So the eigenvalues of this matrix are $\lambda_{1,2} = \pm i \gamma \sqrt{B_x(B_x - B_K)}$, purely imaginary, and are thus center type. Under this condition, there is only a center in the phase space, and the homoclinic orbits disappear as shown in Fig.~\ref{fig:Phase portrait}(c). Therefore, we can expect that the chaotic dynamics induced by the homoclinic orbit will disappear. 

As the external magnetic field increases from $B_x = 0$~mT, the two centers move closer together until they merge into a single center at $(\theta^*, \varphi^*) = (\pi/2,0)$ in the case of $B_x = B_K$ and then there is only one center in the phase space. This is exactly the pitchfork bifurcation. With respect to the magnetic potential, the potential shape changes from a double-well potential to a single-well potential at the pitchfork bifurcation. From the perspective of physics, we can say that the potential shape is important for the generation of chaos.

\section{Numerical analysis}\label{Numerical analysis}
In this section, we demonstrate the emergence of chaos in the presence of two different external periodic forces that drive the magnetization dynamics: the Oersted field and SOT. According to the Poincar\'e-Bendixson theorem \cite{Strogatz}, chaos is prohibited in two-dimensional dynamical systems. In order to observe the chaotic dynamics, we extend Eq.~(\ref{polar_LL}) to three dimensions by incorporating external periodic force and damping force.

\begin{figure}[ht]
\begin{centering}
\includegraphics[width=0.47\textwidth,angle=0]{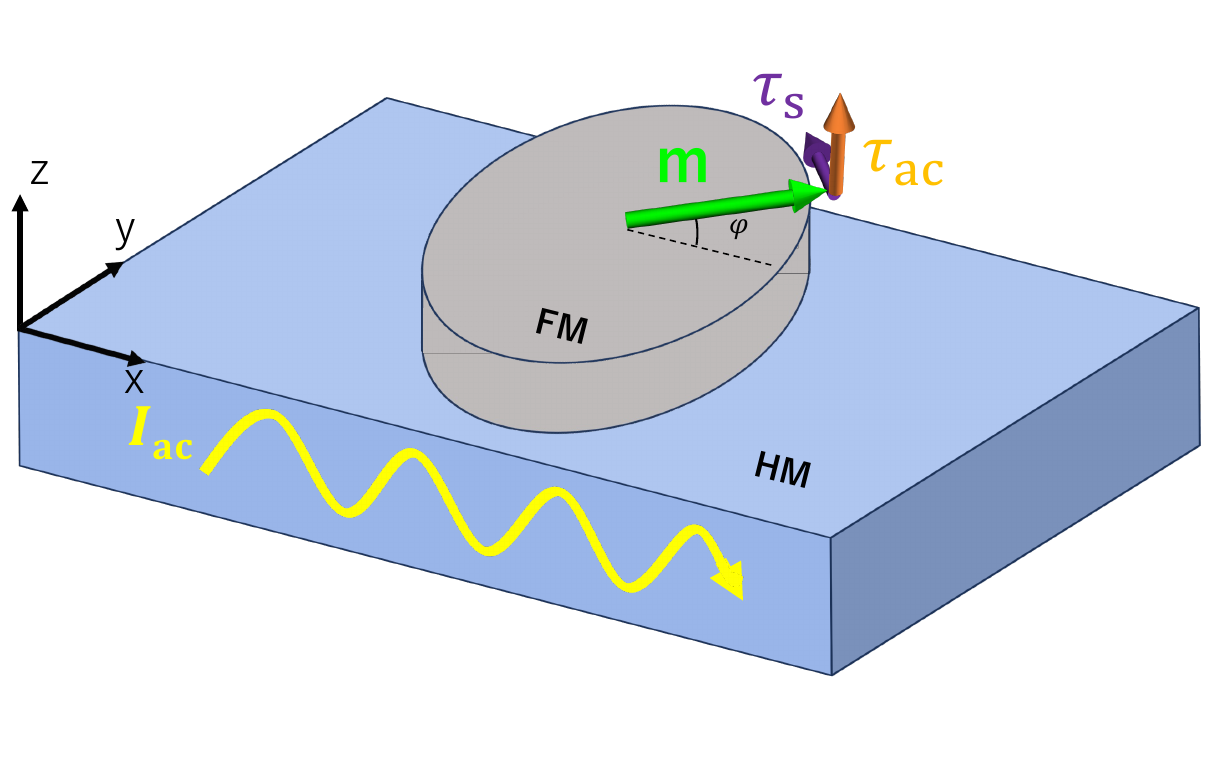} 
\par\end{centering}
\caption{
The illustration depicts a magnetic Duffing oscillator driven by two different external periodic forces: the Oersted field and the spin-orbit torque (SOT) induced by an alternating current $I_{\mathrm{ac}}$ in a heavy metal (HM) layer below the ferromagnet (FM). The Oersted field generates the torque $\bm{\tau}_{\mathrm{ac}}$, which acts in a direction perpendicular to both the $y$ axis and the magnetization. In contrast, the SOT $\bm{\tau}_{\mathrm{s}}$ is a torque that aligns the magnetization with the $y$ axis.
}
\label{fig:devise_3D}
\end{figure}

As shown in Fig.~\ref{fig:devise_3D}, we assume that the alternating current is applied to the heavy metal layer, which can induce an oscillating Oersted's field via Amp\`{e}re's law and an oscillating SOT associated with the spin Hall effect. 
SOT at the ferromagnet/heavy-metal interface can be interpreted as an absorption of spin angular momentum in the heavy metal layer by the ferromagnet \cite{Manchon19}.
The Oersted field generates the following torque:
\begin{equation}
    \bm{\tau}_{\mathrm{ac}}(t)  = - \gamma \bm{m} \times \bm{B}_{\mathrm{ac}}(t),
\label{Oersted_torque}
\end{equation}
where $\bm{B}_{\mathrm{ac}}(t) = B_{\mathrm{ac}} \cos{(\omega_0 t)} \bm{e}_y$ is the Oersted field. 
$\bm{\tau}_{\mathrm{ac}}$ is linear with respect to the magnetization $\bm{m}$ and is directed perpendicular to both the $y$ axis and the magnetization, as shown in Fig.~\ref{fig:devise_3D}. SOT is given by 
\begin{equation}
    \bm{\tau}_{\mathrm{s}}(t) = - \gamma \bm{m} \times (\bm{B}_{\mathrm{s}}(t) \times \bm{m}),
\label{SOT_torque}
\end{equation}
where $\bm{B}_{\mathrm{s}}(t) = B_{\mathrm{s}} \cos{(\omega_0 t)} \bm{e}_y$ is the SOT-induced effective magnetic field. 
$\bm{\tau}_{\mathrm{s}}$ is nonlinear with respected to the magnetization $\bm{m}$ and drives the magnetization to align with the $y$-axis, as shown in Fig.~\ref{fig:devise_3D}.
Here, we assume the resonance condition, i.e., the angular frequency of $\bm{\tau}_{\mathrm{ac}}(t)$ and $\bm{\tau}_{\mathrm{s}}(t)$ is equal to $\omega_0$, which is the resonance (angular) frequency displayed in Fig.~\ref{fig:Phase portrait}.
In order to calculate the magnetization dynamics and the Lyapunov exponent, we represent Eq.~(\ref{LLG equation}) in the polar coordinates as an autonomous system including the Gilbert damping and the external periodic force $\bm{\tau}_{\mathrm{ac}}(t)$ or $\bm{\tau}_{\mathrm{s}}(t)$:
\begin{equation}
\begin{split}
    \frac{d \theta}{dt} &= \gamma B_\varphi(\theta, \varphi) + \alpha \gamma B_\theta(\theta, \varphi) + \tau_\theta(\theta, \varphi, z),\\
    \frac{d \varphi}{dt} &= \frac{1}{\sin{\theta}} \left(  - \gamma B_\theta(\theta, \varphi) + \alpha\gamma B_\varphi(\theta, \varphi) + \tau_\varphi(\theta, \varphi, z)\right),\\
    \frac{d z}{dt} &= \omega_0,
\end{split}
\label{polar_LLG_full}
\end{equation}
where $\tau_\theta$ and $\tau_\varphi$ represent the $\theta$ and $\varphi$ components of the external periodic force: $\bm{\tau}_{\mathrm{ac}}(t)$ or $\bm{\tau}_{\mathrm{s}}(t)$, respectively. 
By solving the differential equation for $z$, we obtain $z = \omega_0 t$. Since cos$(\omega_0 t)$ can be expressed as cos$(z)$, the periodic external forces $\tau_\theta$ and $\tau_\varphi$ in Eq.~(\ref{polar_LLG_full}) depend on $z$.
Throughout this paper, we use the following parameters: $\gamma = 1.76085977 \times 10^{11}$ $\mathrm{T^{-1} s^{-1}}$ and $B_K=200$~mT, whose values can describe the magnetization dynamics of a typical ferromagnet.
Also, we simulated the magnetization dynamics by using the Runge-Kutta method with the DifferentialEquations package in Julia 1.10.
\begin{figure*}[hptb]
\begin{centering}
\includegraphics[width=1\textwidth,angle=0]{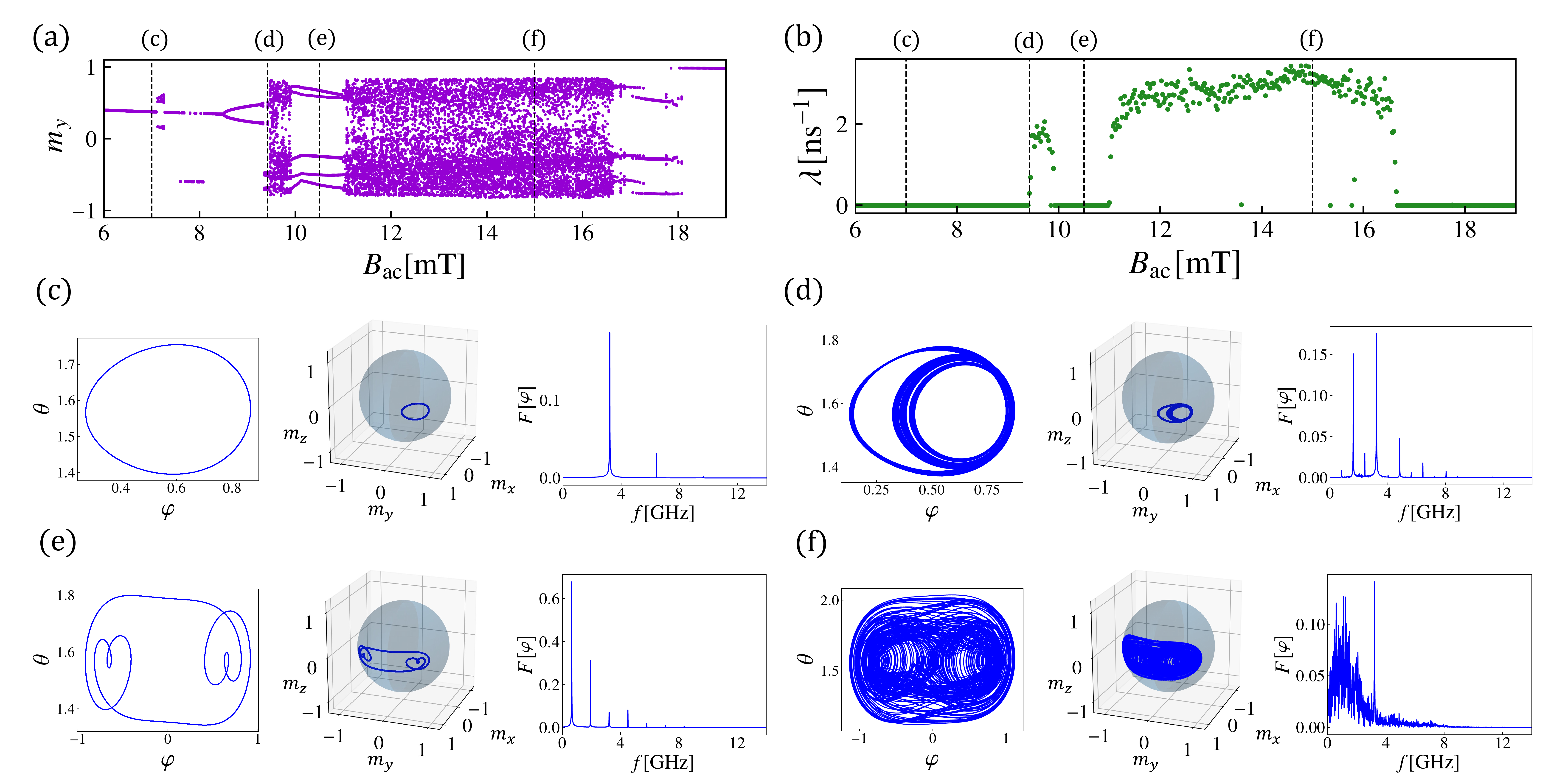} 
\par\end{centering}
\caption{
Oscillatory states of the magnetic Duffing oscillator driven by the Oersted field. (a) The bifurcation diagram as a function of the strength of the Oersted field $B_{\mathrm{ac}}$. (b) The maximum Lyapunov exponent $\lambda$ as a function of the strength of $B_{\mathrm{ac}}$. 
The dynamics is chaotic for $\lambda > 0$ and the dynamics is a limit cycle for $\lambda = 0$. (c) The magnetization dynamics at $B_{\mathrm{ac}} = 7$~mT. The left panel shows a trajectory in the phase space of $\varphi$ and $\theta$. 
The central panel is a trajectory in three dimensions denoted by ($m_x,m_y,m_z$). 
The right panel displays the Fourier spectrum of the azimuth $\varphi$ of magnetization vector $\bm{m}$. (d) The magnetization trajectories and the corresponding Fourier spectrum at $B_{\mathrm{ac}} = 9.425$~mT. The chaotic dynamics trapped within one side of the double-well potential appears. (e) The magnetization trajectories and the corresponding Fourier spectrum at $B_{\mathrm{ac}} = 10.5$~mT. (f) The magnetization trajectories and the corresponding Fourier spectrum at $B_{\mathrm{ac}} = 15$~mT. The chaotic dynamics over the double-well potential appears. For calculating the Lyapunov exponent, the magnetization trajectory, and the Fourier spectrum, we used the magnetization dynamics between $t$ = 700~ns and 800~ns. For the bifurcation diagram, we used the magnetization dynamics between $t$ = 788~ns and 800~ns.
}
\label{fig:Bk=200_mag}
\end{figure*}
\subsection{Bifurcation diagram, Poincar\'{e} section, and Lyapunov exponent}\label{Lya}
In general, chaotic dynamics exhibits nonperiodicity and sensitivity to initial conditions. A bifurcation diagram and Poincar\'{e} section show whether the dynamics is periodic or not. Since the magnetic Duffing oscillator is driven by the oscillating external forces with a time periodicity $T_0$ = $2 \pi / \omega_0$, the system is expected to have a periodic boundary condition of period $T_0$ with respect to time.
Here, $\omega_0$ is the resonance (angular) frequency shown in Fig.~\ref{fig:Phase portrait}. 
Therefore, we make a bifurcation diagram as a stroboscopic map $m_y^* = {m_y(t_0 + nT_0)| n \in \mathbb{N}}$ of the magnetization component $m_y$ after a transient time $t_0$. 
In other words, we plot the value of $m_y$ by time period $T_0$ for each $B_{\mathrm{ac}}$. 
The bifurcation diagram helps us to understand how the periodicity changes as a function of a parameter. 
The Poincar\'{e} section is constructed using a process similar to that of the bifurcation diagram. 
In the proposed model, the Poincar\'{e} section is created by plotting the stroboscopic map, $(\theta, \varphi)^* = {(\theta(t_0 + nT_0), \varphi(t_0 + nT_0)) \mid n \in \mathbb{N}}$, of the state vector in the phase space of $\theta$ and $\varphi$ after a transient time $t_0$. For a periodic state, the section consists of a finite number of points, whereas for quasi-periodic or chaotic states, the Poincar\'{e} section shows densely scattered points. 
Furthermore, in the chaotic state, the Poincar\'{e} section exhibits a fractal-like structure known as the strange attractor.

The Lyapunov exponent is a universal tool for studying nonlinear dynamics and demonstrating the sensitivity to initial conditions.
Since Eq.~(\ref{polar_LLG_full}) is considered as a flow $\dot{\bm{x}} = \bm{f}(\bm{x})$ with $\bm{x} = (\theta,\varphi,z) \in \mathbb{R}^3$, which is a time-dependent vector in three-dimensional phase space, the phase space locally resembles the Euclidean metric and norm $||\bm{x}|| = \sqrt{\Sigma^3_{i=1} x_i^2}$.
The distance between two trajectories, $\bm{x}(t)$ and $\bm{\tilde{x}}(t)$, is defined as $\delta(t) = ||\bm{x}(t) - \bm{\tilde{x}}(t)||$.
The nature of the long-term evolution of $\delta(t)$ for the case $\delta(0) \rightarrow 0$ characterizes the instability of the trajectories. The rate of divergence of the trajectories is determined by the (maximum) Lyapunov exponent
\begin{equation}
    \lambda = \lim_{t \rightarrow \infty} \frac{1}{t} \ln \frac{\delta(t)}{\delta(0)}.
\end{equation}
If $\lambda > 0$, the dynamical system is chaotic, whereas if $\lambda = 0$, the dynamical system is a limit cycle. We calculate the Lyapunov exponent by using the Shimada-Nagashima method \cite{Shimada79}.

\subsection{Chaos driven by Oersted field}\label{with_mag}
To confirm the occurrence of chaos in our model, we begin by analyzing the magnetization dynamics driven by the Oersted field. Hereby, the calculation was performed using only $\bm{\tau}_{\mathrm{ac}}$ as the external periodic force in Eq.~(\ref{polar_LLG_full}).
\begin{figure}[b]
\begin{centering}
\includegraphics[width=0.5\textwidth,angle=0]{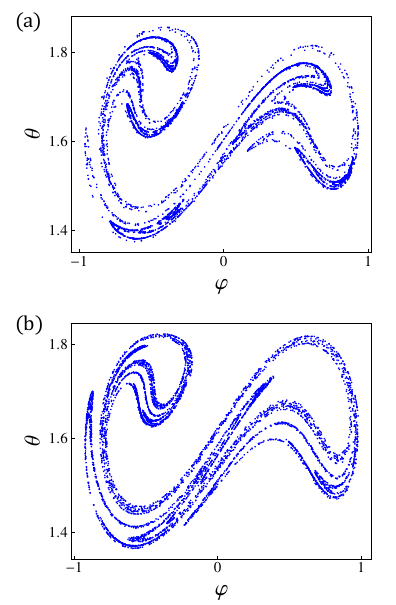} 
\par\end{centering}
\caption{
The Poincar\'{e} section for the magnetic Duffing oscillator only driven by the Oersted field (a) or SOT (b).
}
\label{fig:poincare}
\end{figure}

As seen in Sec.~\ref{Linear_stability}, in the case of $B_x < B_K$, the magnetic Duffing oscillator is expected to exhibit chaos because there are homoclinic orbits in the phase space \cite{Wiggins03}. 
Hence, we use parameters: $B_x = 160$~mT, $B_K = 200$~mT, $\alpha = 0.05$, and $\omega_0/(2\pi)$ = 3.22~GHz. 
Figures~\ref{fig:Bk=200_mag}(a) and (b) illustrate the bifurcation diagram and Lyapunov exponent as a function of the strength of the alternating Oersted field $B_{\mathrm{ac}}$. 
The bifurcation diagram shows the periodicity of the magnetization dynamics for each value of $B_{\mathrm{ac}}$. 
The periodic dynamics is shown in Fig.~\ref{fig:Bk=200_mag}(c). 
The left panel of Fig.~\ref{fig:Bk=200_mag}(c) shows a trajectory of the magnetization dynamics in the phase space of $\theta$ and $\varphi$. 
If the trajectory forms a closed orbit, the same state appears repeatedly at time intervals of $T_0 = 2\pi/\omega_0$. 
The central panel of Fig.~\ref{fig:Bk=200_mag}(c) shows a trajectory of the magnetization dynamics in three dimensions denoted by ($m_x,m_y,m_z$) and a closed orbit.
Although the frequency of the Oersted field is the resonance frequency $\omega_0$, the right panel of Fig.~\ref{fig:Bk=200_mag}(c) shows additional Fourier peaks at the off-resonance frequencies, which indicates that the nonlinear effect appears at $B_{\mathrm{ac}} = 7$~mT.
It is noted that the nonlinear effect manifests as an elliptical trajectory in the phase space of $\theta$ and $\varphi$ and may arise from the Kerr nonlinearity \cite{Zheng23} due to the magnetic anisotropy.
The closed orbit shown in Fig.~\ref{fig:Bk=200_mag}(c) is a limit cycle, as suggested by the bifurcation diagram and the Lyapunov exponent. In contrast, the strong nonlinear dynamics is observed at $B_{\mathrm{ac}} = 9.425$~mT, as shown in Fig.~\ref{fig:Bk=200_mag}(d). 
There are a number of peaks in the Fourier spectrum in the right panel of Fig.~\ref{fig:Bk=200_mag}(c), and the trajectory is complex, indicating that the same state does not appear for a long time. 
According to the Lyapunov exponent $\lambda$ at $B_{\mathrm{ac}} = 9.425$~mT in Fig.~\ref{fig:Bk=200_mag}(a), the dynamics is likely to be chaotic.
As $B_{\mathrm{ac}}$ increases, the magnetization dynamics transition from being confined to one side of the double-well potential to traversing over the double-well potential. In the case of $B_{\mathrm{ac}} = 10.5$~mT, the closed orbit over the double-well potential is observed in Fig.~\ref{fig:Bk=200_mag}(e).
In this condition, the dynamics represents a limit cycle because the trajectory is relatively simple, and the Lyapunov exponent $\lambda$ at $B_{\mathrm{ac}} = 10.5$~mT in Fig.~\ref{fig:Bk=200_mag}(a) is zero.
On the other hand, at $B_{\mathrm{ac}} = 15$~mT, we can observe the chaotic dynamics over a double-well potential, as shown in Fig.~\ref{fig:Bk=200_mag}(f). 
In this case, the Fourier spectrum is broad, and the trajectory covers a wide area in the phase space. 
Figure~\ref{fig:poincare}(a) displays the Poincar\'{e} section of the magnetization dynamics in the phase space of $\theta$ and $\varphi$ at $B_{\mathrm{ac}} = 15$~mT. 
The section shows a fractal-like structure, which resembles that of the Duffing oscillator under chaotic conditions shown in Fig~\ref{fig:Duffing_poincare}.
In addition, the Lyapunov exponent $\lambda$ at $B_{\mathrm{ac}} = 15$~mT is greater than zero. 
For these reasons, we can conclude that chaos occurs in the magnetic Duffing oscillator driven by the Oersted field.
\begin{figure*}[!htpb]
\begin{centering}
\includegraphics[width=1\textwidth,angle=0]{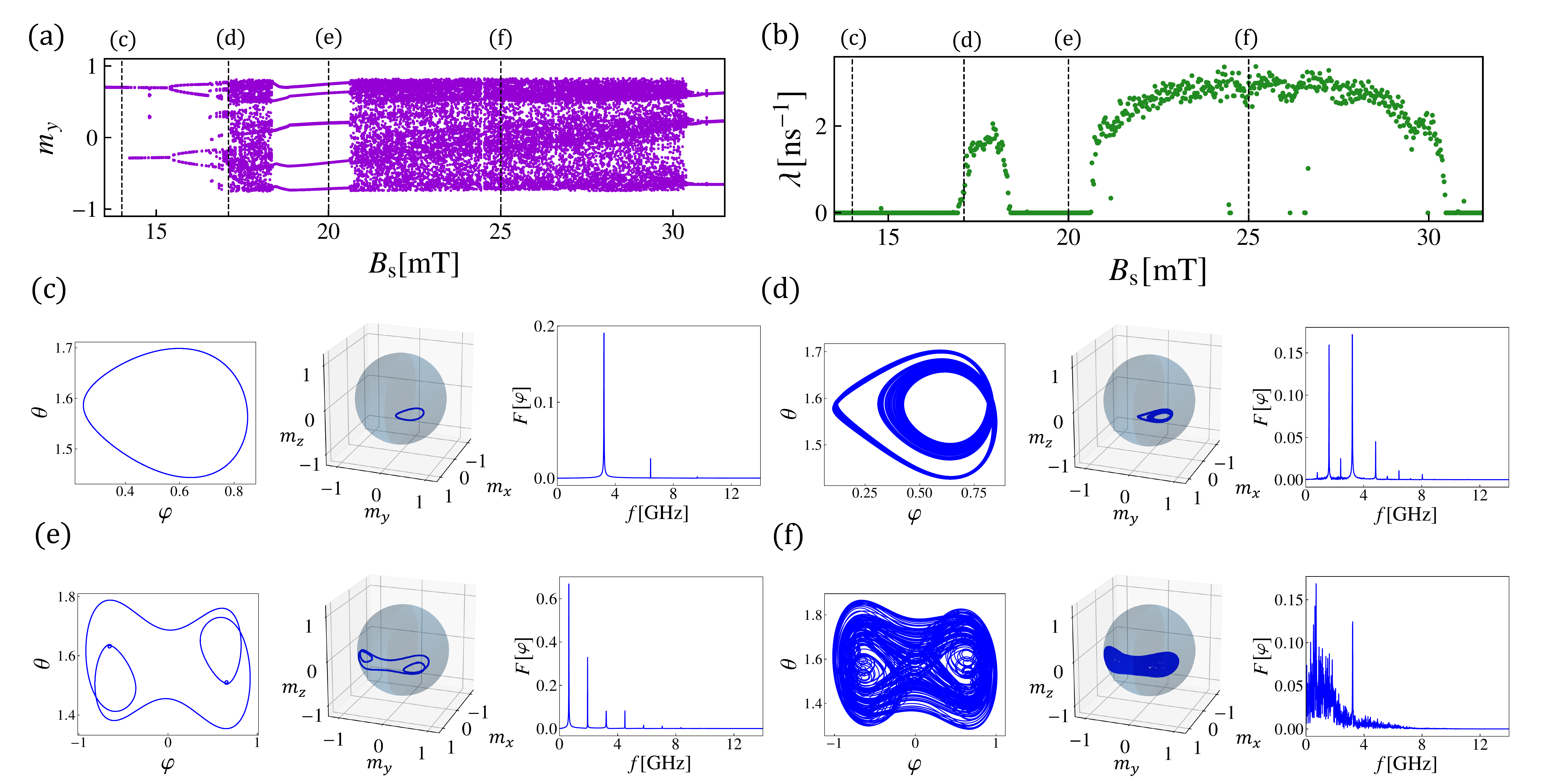} 
\par\end{centering}
\caption{
Oscillatory states of the magnetic Duffing oscillator driven by SOT. (a) The bifurcation diagram as a function of the strength of the SOT-induced magnetic $B_{\mathrm{s}}$. (b) The Lyapunov exponent $\lambda$ as a function of the strength of $B_{\mathrm{s}}$. The dynamics is chaotic for $\lambda > 0$ and the dynamics is a limit cycle for $\lambda = 0$. (c) The magnetization dynamics at $B_{\mathrm{s}} = 14$~mT. The left panel shows a trajectory in the phase space of $\varphi$ and $\theta$. The central panel is a trajectory in three dimensions denoted by ($m_x,m_y,m_z$). The right panel displays the Fourier spectrum with frequency of the azimuth $\varphi$ of magnetization vector $\bm{m}$. (d) The magnetization trajectory and the corresponding Fourier spectrum at $B_{\mathrm{s}} = 17.1$~mT. The chaotic dynamics appears. (e) The magnetization trajectory and the corresponding Fourier spectrum at $B_{\mathrm{s}} = 20$~mT. (f) The magnetization trajectory and the corresponding Fourier spectrum at $B_{\mathrm{s}} = 25$~mT. The chaotic dynamics over double-well potential appears. We used the dynamics between $t$ = 700~ns and 800~ns for calculating the Lyapunov exponent, the magnetization trajectory, and the Fourier spectrum. For the bifurcation diagram, we used the magnetization dynamics between $t$ = 788~ns and 800~ns.
}
\label{fig:Bk=200_spin}
\end{figure*}

According to the bifurcation diagram and the Lyapunov exponent shown in Figs~\ref{fig:Bk=200_mag}(a) and \ref{fig:Bk=200_mag}(b), the periodic dynamics with period $T_0$ occurs at first.~Thus, this is the ferromagnetic resonance. 
Where $B_{\mathrm{ac}}$ is around 8~mT, the bifurcation diagram shows that the magnetization dynamics occasionally shift to $m_y<0$ as $B_{\mathrm{ac}}$ increases. This behavior is due to the instability of the transition process. 
If the magnetization is trapped on the right- (left)-hand side of the potential, where $\varphi>0$ ($\varphi<0$) after the transition process, the periodic dynamics is observed in $m_y>0$ ($m_y<0$) indefinitely. 
Actually, when $B_{\mathrm{ac}}$ is less than 9.425~mT, no points are simultaneously plotted at both $m_y>0$ and $m_y<0$ for the same value of $B_{\mathrm{ac}}$. 
Chaotic dynamics first occurs around $B_{\mathrm{ac}} = 10$~mT. 
Then, a region of the limit cycle, called the periodic window, appears around $B_{\mathrm{ac}} = 11$~mT, after which chaos occurs again.
From the viewpoint of fundamental physics, these trajectory, bifurcation and Poincar\'{e} section are similar to those of the Duffing oscillator [see Appendix~\ref{Duffing equation}]. 

\begin{figure*}[hptb]
\begin{centering}
\includegraphics[width=1\textwidth,angle=0]{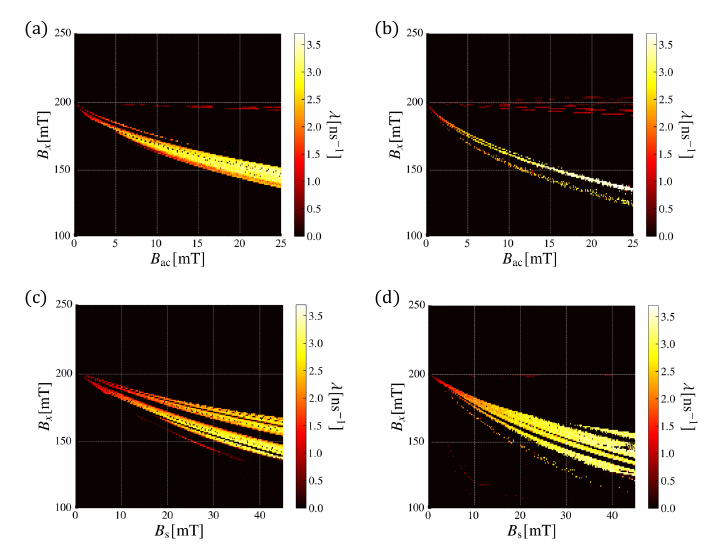} 
\par\end{centering}
\caption{
The Lyapunov exponent $\lambda$ as a function of the static external magnetic field $B_x$ and the strength of the external periodic forces: the Oersted field $B_{\mathrm{ac}}$ for (a) $\alpha=0.05$ and (b) $\alpha = 0.01$, and SOT field $B_{\mathrm{s}}$ for (c) $\alpha=0.05$ and (d) $\alpha = 0.01$. The blight area indicates $\lambda > 0$.}
\label{fig:Lya_maps}
\end{figure*}

\subsection{Chaos driven by spin-orbit torque}\label{With_spin}
Here, we investigate magnetization dynamics driven only by SOT $\bm{\tau}_{\mathrm{s}}$ to confirm the generation of chaotic dynamics other than the Oersted field. 
We consider the same situation $B_x < B_K$ by using parameters: $B_x = 160$~mT, $B_K = 200$~mT, $\alpha = 0.05$, and $\omega_0/(2\pi)$ = 3.22~GHz. 
Figures~\ref{fig:Bk=200_spin}(a) and \ref{fig:Bk=200_spin}(b) are the bifurcation diagram and Lyapunov exponent as a function of the strength of the SOT-induced magnetic field $B_{s}$.
The left and central panels of Fig.~\ref{fig:Bk=200_spin}(c) show a trajectory of the magnetization dynamics and exhibit a closed orbit. 
As in the case driven by the Oersted field, the right panel of Fig.~\ref{fig:Bk=200_spin}(c) shows an additional Fourier peak at the off-resonance frequency, which indicates that the nonlinear effect appears at $B_{\mathrm{s}} = 12$~mT. 
As indicated by the bifurcation diagram and Lyapunov exponent at $B_{\mathrm{s}} = 12$~mT, the limit cycle is observed in Fig.~\ref{fig:Bk=200_spin}~(c).
According to the Lyapunov exponent at $B_{\mathrm{s}} = 16.6$~mT, as in the case of the Oersted field, the chaotic dynamics within one side of the double-well potential is observed at $B_{\mathrm{s}} = 16.6$~mT in Fig.~\ref{fig:Bk=200_spin}(d).  
As the strength of $B_{s}$ increases to 20~mT, we observe periodic dynamics over the double-well potential in Fig.~\ref{fig:Bk=200_spin}(e).
In this condition, there are some Fourier peaks and the trajectory is not complicated. Then, the Lyapunov exponent at $B_{\mathrm{s}} = 20$~mT is zero.
At $B_{\mathrm{s}} = 25$~mT, chaotic dynamics is observed over a double-well potential in Fig.~\ref{fig:Bk=200_spin}(f).
Here, the Fourier spectrum is broad and the trajectory covers a wide area in the phase space. The Poincar\'{e} section shows a fractal-like structure in Fig.~\ref{fig:poincare}(b) and the Lyapunov exponent is greater than zero at $B_{\mathrm{s}} = 25$~mT. 

As in Sec.~\ref{with_mag}, an increase in the external periodic force induced a transition from chaotic dynamics trapped within one side of the double-well potential to chaotic dynamics traveling over the double-well potential. 
As shown in Appendix \ref{Duffing}, this behavior is observed in the Duffing oscillator.
The Poincar\'{e} section in the case of the Oersted field and those associated with SOT exhibit similar characters.
The generation of the chaotic magnetization dynamics driven by either the Oersted field or SOT has been confirmed. 
However, there are some quantitative differences: the parameter ranges for the SOT-driven chaos are wider than those for the Oersted field, as seen in Figs.~\ref{fig:Bk=200_mag} and~\ref{fig:Bk=200_spin}.
The Oersted field produces a torque that induces the magnetization to move along an isoenergetic curve in the phase space, whereas the SOT is a torque that displaces the magnetization along the gradient of the energy surface.
As discussed in Sec.~\ref{Linear_stability}, the homoclinic orbit is one of the isoenergetic curves. 
Therefore, the torque direction changes the parameter ranges that generate chaotic dynamics.

\subsection{Correlation between chaos and homoclinic orbit}\label{Lya_maps}
Here, we demonstrate that the chaos observed in the magnetic Duffing oscillator originates from the homoclinic orbit in the phase space. To this end, we investigate the influence of deformation of the magnetic potential by the static external magnetic field. 
Especially, as discussed in Sec.~\ref{Linear_stability}, chaotic dynamics is expected to disappear when the external magnetic field $B_x$ exceeds $B_K = 200$~mT because the homoclinic orbits vanish from the phase space after the pitchfork bifurcation occurs.
For the investigation, we use $B_K = 200$~mT and the resonance frequency $\omega_0/(2\pi)$ is calculated by the eigenvalues of the Jacobian matrix at the center point. 

Figure~\ref{fig:Lya_maps}(a) shows the Lyapunov exponent as functions of the static external magnetic field $B_x$ and the Oersted field $B_{\mathrm{ac}}$ for $\alpha = 0.05$. In the context of $\lambda > 0$, the blight areas in Fig.~\ref{fig:Lya_maps}(a) indicate chaotic magnetization dynamics. 
When the external magnetic field $B_x$ decreases from 200~mT, the threshold strength of the Oersted field required for the onset of chaotic behavior increases, and the range of the Oersted field values that generate chaotic dynamics is also expanded.
As the value of $B_x$ approaches 200~mT, we observe that the magnitude of the Oersted field required to generate chaotic behavior approaches zero. 
In particular, the strength of the Oersted field that generates chaotic dynamics at $B_x = 195$~mT is 1~mT, which is two orders smaller than the magnetic anisotropy field $B_K$. 
The external magnetic field plays a role in reducing the potential barrier of the double-well potential and making it easier to pass through the neighborhood of the homoclinic orbits. 
This indicates that the position of the homoclinic orbit can be modified by an external magnetic field, which can adjust the parameter range where chaos occurs.
Chaotic dynamics disappears when $B_x$ exceeds 200~mT, since the homoclinic orbits disappear above $B_x = 200$~mT, as discussed in Sec.~\ref{Linear_stability}. 

Figure~\ref{fig:Lya_maps}(b) shows the calculated Lyapunov exponent for $\alpha = 0.01$. It is found that the smaller value of $\alpha$ reduces the strength of the Oersted field required to induce chaotic dynamics. 
However, the qualitative behavior remains the same for both cases of $\alpha = 0.01$ and $0.05$ before the pitchfork bifurcation occurs. 
There are some regions where the Lyapunov exponent is greater than zero even when $B_x$ exceeds 200~mT for $\alpha = 0.01$, whose behavior is not observed for $\alpha = 0.05$ shown in Fig.~\ref{fig:Lya_maps}(a). 
This chaotic dynamics occurs near the pitchfork bifurcation ($B_x \sim B_K$), i.e., in regions where the geometric structure of the vector field in the phase space is highly sensitive to perturbations such as external periodic forces and damping force. 
Since these forces are not taken into account in the linear stability analysis, the chaotic behavior that is not predicted by the linear stability analysis emerges. 
Thus, although the general picture of the dynamics can be predicted by the linear stability analysis, unexpected chaotic dynamics occurs near the critical point, the pitchfork bifurcation.

Next, the Lyapunov exponent as functions of the static external magnetic field $B_x$ and SOT-induced magnetic field $B_{\mathrm{s}}$ for $\alpha = 0.05$ and $\alpha = 0.01$ are shown in Figs.~\ref{fig:Lya_maps}(c) and~\ref{fig:Lya_maps}(d).
Similar to Figs.~\ref{fig:Lya_maps}(a) and~\ref{fig:Lya_maps}(b), before the pitchfork bifurcation, the threshold strength of the SOT required for the onset of chaotic behavior increases, and the range of the SOT values that generate chaotic dynamics is also expanded.
Chaotic dynamics cannot be found when $B_x$ exceeds 200~mT for both  $\alpha = 0.05$ and $\alpha = 0.01$.
In contrast to the case of the Oersted field, the range of the SOT-induced magnetic field that generates chaotic dynamics does not significantly change when the Gilbert damping constant $\alpha$ is 0.01 or 0.05.
It can also be understood by considering the direction of the torque.
Gilbert damping produces a torque that aligns the magnetization along the direction of the gradient of the energy surface, whereas the Oersted field produces a torque along the isoenergetic curve.
At steady state, the magnetization rotates in such a way that the energy gained from the AC input is balanced by the energy loss due to damping. Chaos arises when energy loss and gain are balanced near the homoclinic orbit, which is one of the isoenergetic curves. 
Consequently, when the damping constant $\alpha$ is small, a slight increase of the Oersted field causes the magnetization trajectory to deviate from the homoclinic orbit, thereby narrowing the parameter range of chaotic behavior.
In contrast, SOT is a torque that drives the magnetization up and down along the gradient of the energy surface, resulting in the magnetization trajectory out of the isoenergetic curves. 
Therefore, the dependence of the Lyapunov exponent on $\alpha$ differed between the Oersted field and SOT.

\section{Discussion}
\label{Discussion}
In this study, we explained one of the mechanisms of chaos in a spintronics system driven by external periodic forces, where a uniaxial magnetic anisotropy and an external magnetic field are an orthogonal configuration. 
This circumstance can be realized in various spintoronics devices. 
In the first part of this section, we discuss the chaotic magnetization dynamics induced by the homoclinic orbit in one of the spintronics devices: the nano-oscillator driven by the VCMA effect.
The chaos described in Ref.~\onlinecite{Contreras-Celada22} occurs under the condition where a uniaxial magnetic anisotropy and an external magnetic field are the canted configuration. 
By utilizing the VCMA effect, the magnetic anisotropy field is changed by AC voltage at a metal-insulator interface. 
When the chaotic magnetization dynamics occurs, the homoclinic orbit should appear due to the change in the magnetic anisotropy.
This can be regarded as a situation in which an external periodic force is applied to a system with the homoclinic orbit in the phase space, such as the magnetic Duffing oscillator. 

Also, our study suggests that chaotic magnetization dynamics may occur within a realistic parameter range in the ST-FMR \cite{Liu11,Chiba14} which is a widely used technique for investigating spintronic phenomena, particularly in systems with a strong spin-orbit coupling. This method involves applying an alternating current to a bilayer film composed of a ferromagnet, such as permalloy, and a heavy metal, such as platinum. The alternating current generates both the Oersted field and SOT on the ferromagnet. Consequently, magnetic precession occurs, which can be detected by a resultant voltage signal across the bilayer film. Then, the information of chaotic magnetization dynamics is embedded in the voltage signal.

Finally, we introduce some studies that approximate the LLG equation under certain assumptions and express it in a form similar to the Duffing equation. 
Ref.~\onlinecite{Shukrinov21} investigates nonlinear magnetization dynamics in interacting superconductor-ferromagnet systems, and Ref.~\onlinecite{Moon14} examines the chaotic dynamics of magnetic vortex states in a circular nanodisk.
Since these studies demonstrate the correspondence between the Duffing equation and the LLG equation by expanding the LLG equation in the vicinity of a particular point in the phase space, it is difficult to determine whether this correspondence holds globally across the phase space.
In contrast, the qualitative behavior of the magnetic Duffing oscillator and the Duffing oscillator remains consistent well beyond the equilibrium point—a fact that cannot be easily captured by local analyses such as a Taylor expansion.
The Duffing equation is formulated in the Euclidean coordinate system, which has no curvature, whereas the LLG equation describes dynamics on the curved phase space $S^2$.
This difference prevents a global correspondence between the equations' forms, even though they can exhibit the same qualitative behavior.
However, geometric methods such as phase portraits and Poincar\'e sections, which have been developed to analyze dynamical systems, have successfully addressed the above difficulties.
Thus, we believe that employing geometric methods is a reasonable and effective approach for studying magnetization dynamics, which exhibits rich nonlinear effects in curved space.

\section{Conclusion}
\label{Conclusion}
In conclusion, we proposed the magnetic Duffing oscillator that exhibits chaotic magnetization dynamics due to the double-well magnetic potential and demonstrates similar behavior to that of the Duffing oscillator. 
Based on the linear stability analysis of the LLG equation, we found that an external magnetic field applied perpendicular to the magnetic anisotropy field creates an anharmonicity on the magnetic potential, generating homoclinic orbits in the phase space. 
In the case of $B_x < B_K$, the phase portrait of the magnetic Duffing oscillator exhibits the same topology as that of the Duffing oscillator, implying that both systems exhibit qualitatively the same behavior.
Based on the Lyapunov exponent, the chaotic dynamics driven by the Oersted field and SOT are confirmed. 
This result indicates that the magnetic Duffing oscillator does not need to assume some specific driving force to generate chaotic dynamics.
As seen in the Duffing oscillator, chaotic dynamics confined within one side of the double-well potential and traversing over the double-well potential are observed. 
Furthermore, we confirmed that the trajectory and Poincar\'{e} section in the phase space of $\varphi$ and $\theta$ show a behavior similar to that of the Duffing oscillator. 
We also calculated the Lyapunov exponent as a function of an external magnetic field for different relaxation strengths of $\alpha$.
Theoretical predictions of the parameter range exhibiting chaotic dynamics were made through an analysis of the phase portrait and were confirmed numerically through the simulations. 
Indeed, while the theoretical analysis provides a general prediction of the magnetic Duffing oscillator behavior, unexpected chaotic dynamics also emerges. 
In addition, we demonstrated that the regions of chaos are controlled by adjusting an external magnetic field, which is of practical importance for designing spintronics devices that utilize nonlinear and chaotic magnetization dynamics.
We further discussed the applicability of the presented model to chaos reported so far in spintronic devices and proposed how to realize the magnetic Duffing oscillator in an experiment.
This work enables us not only to enhance our understanding of the chaotic dynamics of the magnetization but also to aid in the design of spintronic devices that exhibit chaotic magnetization dynamics.

\section*{Acknowledgments}
The authors thank Tomohiro Taniguchi, Shinji Miwa, Sadamichi Maekawa, and Yusuke Hashimoto for valuable discussions. This work was supported by Grants-in-Aid for Scientific Research (Grants No.~22K14591 and No.~24K00916). HM acknowledges support from CSIS, Tohoku University.

\appendix

\section{Duffing oscillator}\label{Duffing}
In order to show a typical dynamics with the homoclinic orbital, we introduce the Duffing equation in Eq.~(\ref{Duffing equation}), which is a famous equation exhibiting chaotic dynamics in the Hamiltonian systems
\begin{equation}
    \frac{d^2 x}{dt^2} + \alpha \frac{dx}{dt} + \beta x + \gamma x^3 = F_0 \sin{\mathrm{\omega_0 t}},
    \label{Duffing equation}
\end{equation}
where, $\alpha$ and $F_0 \sin{\mathrm{\omega_0 t}}$ are the damping coefficient and the external periodic force with frequency $\omega_0$.
\begin{figure}[h]
\includegraphics[width=0.48\textwidth,angle=0]{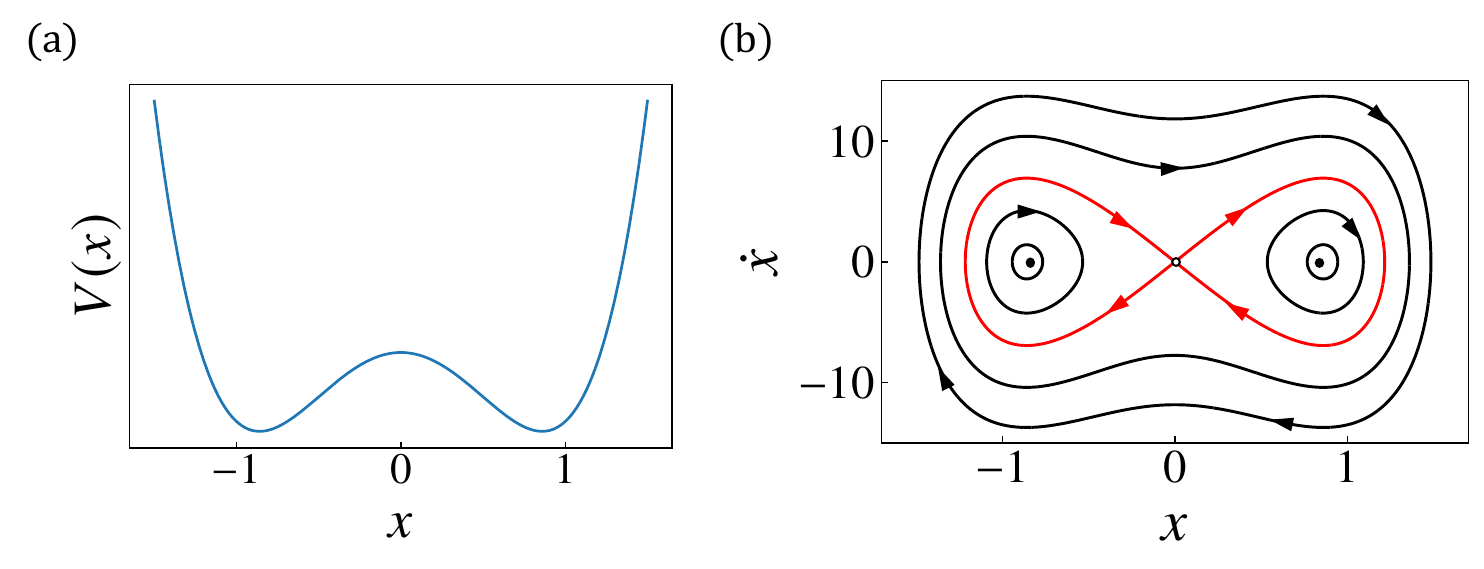} 
\caption{
(a) The potential of the Duffing equation for $\beta<0$, where $\beta = - 130, \gamma = 176$. The shape of the potential is a double-well potential and chaotic dynamics appear. (b) The Phase portraits of the Duffing oscillator excluding relaxation and external periodic force. A filled circle represents the center and an open circle represents the saddle. The red line is the homoclinic orbit, which
is the trajectories that start from and end at the saddle point.}
\label{fig:Duffing_potential}
\end{figure}
\begin{figure*}[ht]
\includegraphics[width=1\textwidth,angle=0]{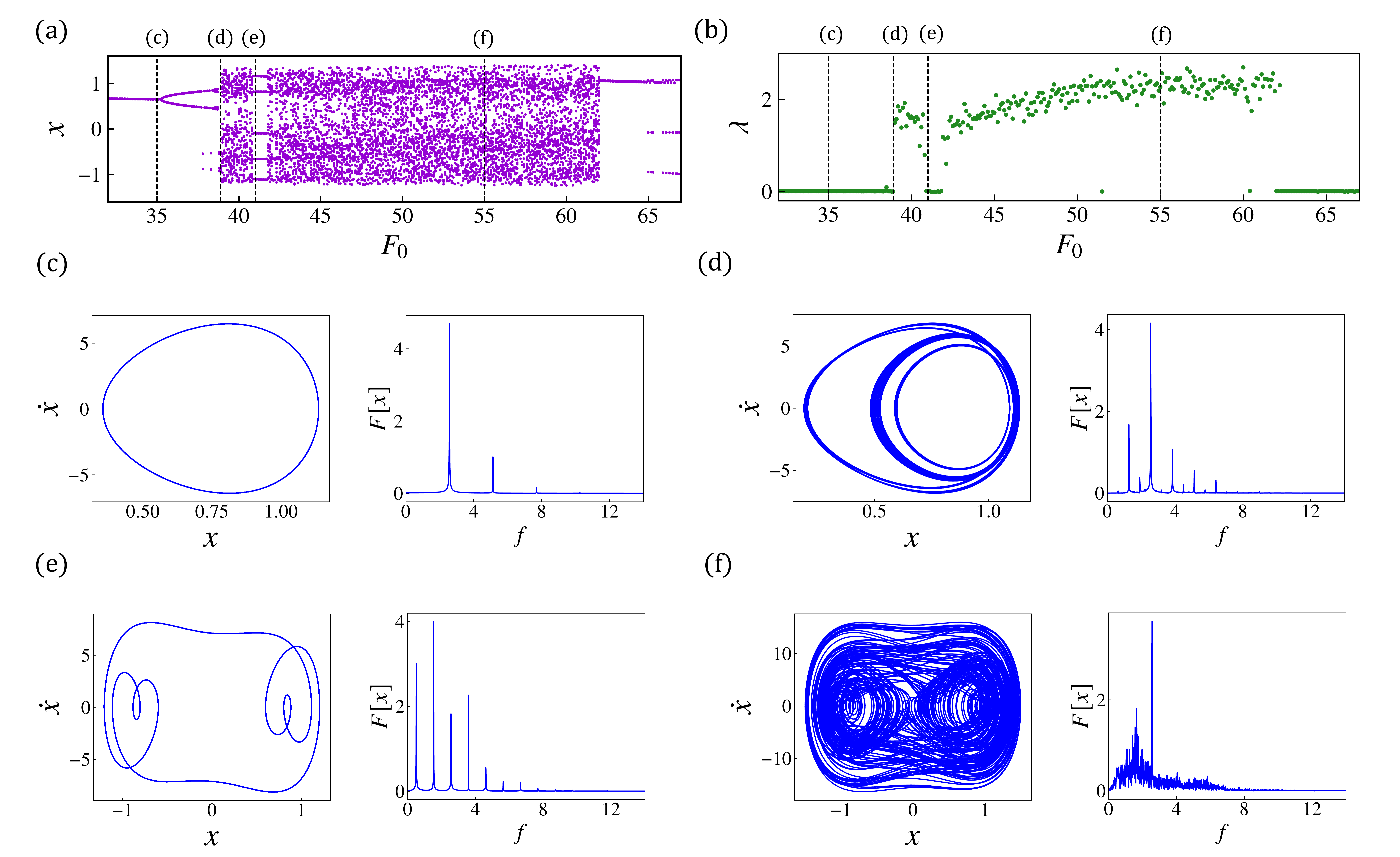} 
\caption{
Oscillatory states of the Duffing oscillator. (a) The bifurcation diagram as a function of the strength of the external periodic force $F_0$. (b) The Lyapunov exponent $\lambda$ as a function of the strength of the external periodic force $F_0$. (c) The mass point dynamics at $F_0 = 35$. The left panel shows a trajectory in the phase space of $x$ and $\dot{x}$. The right panel displays the Fourier spectrum with frequency of the mass position $x$. (d) The mass point trajectory and the corresponding Fourier spectrum at $F_0 = 38.9$. The chaotic dynamics appears. (e) The mass point trajectory and the corresponding Fourier spectrum at $F_0 = 41$. (f) The mass point trajectory and the corresponding Fourier spectrum at $F_0 = 55$. The chaotic dynamics over a double-well potential appears. We used the dynamics between $t$ = 700 and 800 for calculating the Lyapunov exponent, trajectory, and the Fourier spectrum. Additionally, we used the dynamics between $t$ = 788 and 800 for the bifurcation diagram.
}
\label{fig:Duffing_result}
\end{figure*}
\begin{figure}[h]
\begin{centering}
\includegraphics[width=0.5\textwidth,angle=0]{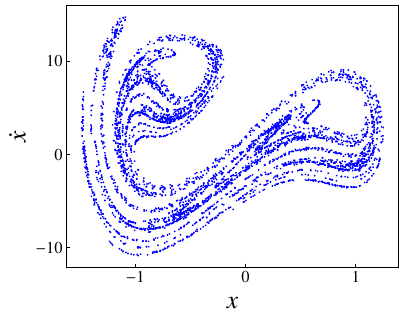} 
\par\end{centering}
\caption{
Poincar\'{e} section for the Duffing oscillator. The values of the parameter are the same as in Fig.~\ref{fig:Duffing_result}~(f).
}
\label{fig:Duffing_poincare}
\end{figure}
The Duffing equation describes the mass point dynamics within the potential $V(x) = \beta x^2/2 + \gamma x^4 /4$. 
For $\beta < 0$, the shape of the potential is a double-well potential [see Fig.~\ref{fig:Duffing_potential}(a)] and the phase space has the homoclinic orbit as shown in Fig.~\ref{fig:Duffing_potential}(b). 
Therefore, the chaotic behavior appears in the Duffing equation. 

Figures~\ref{fig:Duffing_result}(a) and \ref{fig:Duffing_result}(b) show the Lyapunov exponent and bifurcation diagram as a function of the external periodic force $F_0$. The parameters are $\alpha = 1, \beta = -130, \gamma = 176, \omega_0 / (2\pi) = 2.56$. 
The left and central panels of Fig.~\ref{fig:Duffing_result}(c) have a closed orbit, and the right panel of Fig.~\ref{fig:Duffing_result}(c) has a Fourier peak at $\omega/(2\pi)\neq \omega_0/(2\pi)$.
Thus, the nonlinear effect appears at $F_0 = 25$. It is a limit cycle, as suggested by the bifurcation diagram and Lyapunov exponent. Chaotic dynamics confined within one side of the double-well potential is shown at $F_0 = 38.66$ in Fig.~\ref{fig:Duffing_result}(d). There are numerous Fourier spectrum peaks, and the trajectory is complex. According to the Lyapunov exponent, the dynamics seems to be chaotic. Increasing the strength of the periodic force to $41$, we observe the periodic dynamics between a double-well potential in Fig.~\ref{fig:Duffing_result}(e). 
Under this condition, there are some Fourier spectra, but the trajectory is closed, and the Lyapunov exponent is less than zero. Finally, we can observe the chaotic dynamics over a double-well potential in Fig.~\ref{fig:Bk=200_mag}(f) at $F_0 = 50$. Under this condition, the Fourier spectrum is broad, and the trajectory covers a wide area in the phase space. The Lyapunov exponent is greater than zero. The trajectory and the Poincar\'{e} section in the phase space of $x$ and $\dot{x}$ are similar to those of the magnetic Duffing oscillator, as shown in Secs.~\ref{with_mag} and \ref{With_spin}. The shape of the trajectory, such as the heart shape in Fig.~\ref{fig:Duffing_result}(e), and the transition of dynamics from being confined within one side of the double-well potential to traversing over the double-well potential are exhibited, as is the case with the magnetic Duffing oscillator.
Figure~\ref{fig:Duffing_poincare} displays the Poincar\'{e} section of the Duffing oscillator in the phase space of $x$ and $\dot{x}$ at $F_0 = 55$. 
The section shows a fractal-like structure.

\end{document}